# Optical phase locking of two infrared CW lasers separated by 100 THz

N. Chiodo[1], F. Du-Burck[2], J. Hrabina[3], M. Lours[1], E. Chea[1] and O. Acef [1*]

[1] LNE-SYRTE, Observatoire de Paris/CNRS-UMR 8630/UPMC Paris VI, 61 avenue de l'Observatoire, 75014 Paris, France
[2] LPL/ CNRS-UMR 7538 / Université Paris 13 - Sorbonne Paris Cité, 99 avenue J. B. Clément, 93430 Villetaneuse, France
[3] Institute of Scientific Instruments of the ASCR, v.v.i., Královopolská 147, 61264 Brno, Czech Republic
*Corresponding author: ouali.acef@obspm.fr

We report on phase-locking of two continuous wave infrared laser sources separated by 100 THz emitting around 1029 nm and 1544 nm respectively. Our approach uses three independent harmonic generation processes of the IR laser frequencies in periodically poled MgO: LiNbO$_3$ crystals to generate second and third harmonic of that two IR sources. The beat note between the two independent green radiations generated around 515 nm is used to phase-lock one IR laser to the other, with tunable radio frequency offset. In this way, the whole setup operates as a mini frequency comb (MFC) emitting four intense optical radiations (1544 nm, 1029 nm, 772 nm and 515 nm), with output powers at least 3 orders of magnitude higher than the available power from each mode emitted by femtosecond lasers.

Powerful and phase coherent multi-radiations are important in various domains of metrology and fundamental physics and are already involved in various experiments such as broadband free space optical telecommunications, accurate distance measurements through atmosphere, determination of associated atmospheric phase shift, up/down optical transmissions between Earth and satellites, etc [1]. Nowadays, harmonic generation processes and frequency mixing allowed by the recent development of very high efficient non-linear crystals, offer the possibility to connect radiations separated by large frequency gaps in the optical domain, and to phase lock them to each other [2,3].
We report in this paper the phase locking of two continuous wave (cw) infrared laser (IR) sources separated by a frequency difference of 100 THz, using frequency doubling and tripling processes, which allow generation of a quadruplet of phase coherent optical radiations in both IR and visible domains. The available optical powers range from few milliwatts to hundreds of milliwatts, largely higher than the available power from each mode of a frequency comb with similar frequency span [4,5]. The green radiation generated by our setup is in a spectral region where molecular iodine shows lots of narrow (a few hundreds of kilohertz) and intense absorption lines [6]. The use of one of them as frequency reference will give the possibility to transfer accuracy and stability to all four generated radiations, in particular to the telecom laser at 1544 nm. A relative stability in the range $10^{-13}$ -$10^{-14}$ at 1 s integration time is expected [7]. This may be compared to telecom laser directly locked to $C_2H_2$ transitions around 1500 nm, for which stabilities in the range $10^{-11}$-$10^{-13}$ are reported [8,9].
The optical setup is depicted in figure 1. The first IR laser source is an Yb-doped fibre laser operating around 1029 nm, followed by an optical amplifier emitting more than 0.5 W of output power. The fundamental wavelength emitted by this laser can be tuned from 1028.7 nm to 1029.4 nm by means of temperature control of the Bragg grating and with a ceramic piezo-electric actuator (PZT). The laser linewidth is smaller than 10 kHz before any electronic feedback. The laser system is placed in a metallic box, surrounded by an additional wooden box to ensure thermo-acoustic isolation. The second harmonic generation (SHG) process is operated in a free space waveguide MgO doped periodically poled Lithium Niobate (PPLN) crystal of 10 mm length. This frequency doubling is described in more details in [10]. The quasi phase matching (QPM) condition is obtained by inserting the crystal in a home made temperature controlled oven made of copper. Temperature stability better than 5 mK is currently achieved over more than 10 hours of continuous work. The incident IR beam is coupled into the waveguide crystal by means of a lens of 8 mm focal length (lens L in Fig. 1). A dichroic mirror placed at the output of this crystal separates the fundamental and harmonic beams. More than 10 mW of green radiation is currently obtained day to day, in very stable operational mode using 100 mW of IR incident power (Fig. 2). The wavelength and temperature acceptances were measured as 0.2 nm and 2.4°C respectively.
The second IR source is an external-cavity diode laser (ECDL) tunable from 1500 nm to 1600 nm which delivers about 20 mW over the full optical telecommunication C band. The wavelength is tunable by means of a rotating optical grating driven by an electrical motor. The linewidth of this laser is about 150 kHz due to a large sensitivity to acoustical perturbations associated to the mechanical mounting of the grating. The laser current can be externally driven to compensate the fast frequency variations and is used for the phase lock purpose against the 1029 nm laser. This laser is followed by an Er-doped fibre amplifier (EDFA) delivering up to 600 mW.
The third harmonic generation (THG) of this source is based on the use of two cascaded PPLN bulk crystals in single pass configuration. It should be noted that only few experimental demonstrations of frequency tripling of cw sources emitting in the optical telecommunications C-band near 1550 nm have been reported in the literature. Due to the generally small values of third order non linear optical coefficients, THG is obtained from two second order non linear processes in series, SHG ($\omega + \omega \rightarrow 2\omega$) followed by sum-frequency generation (SFG) ($\omega + 2\omega \rightarrow 3\omega$). For this purpose, one can use consecutive interactions in a higher QPM order PP crystal to phase match both SHG and SFG processes by a single period structure as has been done with pulsed sources [11]. However, attempts with cw sources have not demonstrated a sufficient power of the green radiation to probe the iodine vapour in a cell [12,13]. One can also achieve the THG by cascading two PP structures, optimised for QPM of SHG and SFG respectively as has been implemented for UV generation from a cw source at 1064 nm with two consecutive gratings realized in one crystal [14] or with two separated crystals [15,16]. This 2-step configuration was more recently demonstrated for the tripling of 1565 nm radiation [17].

In the present work, we use a similar approach. THG of the 1544 nm source is based on the use of two cascaded PPLN bulk crystals allowing, first a SHG of the IR beam, followed by a SFG of the two previous beams emerging from the first crystal and both focused in the second one. Both crystals have 50 mm length and 0.5 mm height. The first one is AR–coated for $\omega$ and $2\omega$ wavelengths on both facets. The second one is coated for the three wavelengths ($\omega$, $2\omega$ and $3\omega$) on both facets also. The width is 1 mm for the SHG crystal and 3 mm for the SFG crystal. The grating period of the first crystal is 18.8 µm, while that of the second one is 6.92 µm. The QPM temperature of the SHG and SFG crystals were found to be 44.7°C and 19.2°C respectively, and kept at those values within 1 mK over several hours of continuous work.

The polarisation of the beam at the input of the SHG crystal is controlled by a zero order half-wavelength plate and is focused in this crystal using a lens with focal length of 50 mm ($L_1$ in Fig. 1). For a pump power of 630 mW, a residual IR power of 455 mW is measured at the output of this first PPLN crystal. The two beams that emerge from this first crystal are then directly refocused in the second PPLN crystal with a lens of focal length of 75 mm ($L_2$ in Fig. 1). The lens is AR-coated for the second harmonic wavelength ($2\omega$), but introduces up to 16 % of optical losses for the fundamental wave ($\omega$). At the output of the second crystal (SFG) we obtain 1.5 mW at $3\omega$, 8 mW at $2\omega$ and more than 250 mW at the fundamental radiation ($\omega$).

The red solid curve in Fig. 3 shows the SHG power measured at the output of the SHG crystal. The normalized efficiency is found to be 0.6 %.W$^{-1}$.cm$^{-1}$, a value consistent with those found in literature [18, 19] and a maximum SHG power of 12 mW is obtained from an IR laser pump power of 630 mW.

We have also observed in the same time a weak THG radiation (~100 nW at 515 nm) at the output of this first crystal in addition to the SHG process, probably due to the non phase-matched sum between the fundamental wave and the SHG radiation.

The green dashed curve in Fig. 3 gives the harmonic power relative to the green radiation at 515 nm, measured at the output of the SFG crystal. The global IR to green conversion efficiency is estimated to be about 0.6 %.W$^{-2}$. Thus, we obtained a maximum green power of 1.5 mW at 515 nm for 630 mW of IR pump at 1544 nm. Taking into account the various optical losses involved in the coupling of both wavelengths in the SFG crystal, we estimate the normalized efficiency as 6.8 % W$^{-1}$ cm$^{-1}$. This value is 2 times higher than the one obtained for yellow light generation in [20].

We emphasize the level of the green power generated at 515 nm, high enough to probe directly iodine hyperfine transition using saturated absorption technique. It is worth noting that the iodine lines detection, using saturated absorption technique, requires an optical power of few hundreds of µW only. Using this third harmonic process, it would be possible to transfer high frequency stability using iodine line as references to the C band of optical communications.

It must be noted that the balance of powers shows some losses of IR power both in the SHG and SFG processes. Our non linear crystals of 0.5 mm high and 50 mm long are inserted in 75 mm long ovens with small apertures for the beams. The optimization of the harmonic power needs to strongly focus the beams in the crystals which leads to the partial occultation of the beams at the crystal entrances. This effect introduces 20% and 25% of losses of IR power for SHG and SFG processes respectively. Although this situation is not entirely satisfactory, it was in our case the only way to optimize the harmonic power and to reach 1.5 mW of green power.

The 1544 nm laser is tunable over 100 nm and four grating periods are available in the SHG crystal (18.4, 18.6, 18.8, and 19.0 µm). The generated green wavelength with THG process is actually limited by the QPM temperature achieved by the oven used around the SFG crystal between 514 nm to 518 nm, corresponding to IR wavelength ranging from 1542 nm to 1555 nm. We have characterized the wavelength and the temperature acceptance of the THG process near the middle of this interval using a fundamental radiation at 1549 nm. In this case, we have used the grating period of 19.0 µm for the SHG process with a QPM temperature of 35.1°C. The QPM temperature for the SFG was found to be 70.96°C. Wavelength and temperature acceptance are found do be 0.067 nm and 1.5°C respectively (Fig. 4).

The two green radiations generated independently by frequency doubling of the 1029 nm fibre laser on one hand, and by frequency tripling of the 1544 nm ECDL on other hand, permit to achieve an optical beatnote around 515 nm in a fast silicon photodiode (Fig. 5). This beatnote was mixed with a radio frequency reference and was subsequently used to phase lock the 1544 nm ECDL (which exhibits much higher frequency noise), against the fibre laser. Thus, we realize a phase coherent mini frequency comb (MFC) constituted by four optical radiations in the visible (515 nm & 772 nm) and in the IR (1029 nm & 1544 nm). It is worth noting that the power of each tooth of this MFC is at least 3 orders of magnitude higher than the available power from each mode of a frequency comb.

The green radiation obtained by frequency doubling of the 1029 nm laser will be used to detect narrow and intense hyperfine iodine lines at 515 nm, utilizing the well known saturated absorption technique. In this way, this signal will permit to frequency stabilize the whole optical setup. This part of work devoted to spectroscopy and frequency stabilisation using iodine lines will be subject of another paper.

The two 515 nm harmonic beams issued from the two IR sources are combined by an appropriate beam splitter and superimposed on a fast silicon photodiode (Fig. 5). The incident powers are 0.25 mW from master laser SHG and 0.5 mW from slave laser THG. A half wavelength plate is used to optimize the detected optical beatnote in the Si photodiode. A first part of this beatnote is monitored by a spectrum analyzer, while the second one is frequency divided by a factor 8 before being mixed with the RF reference.

The DC signal from the RF mixer is low pass filtered, and applied to a proportional - integrator (PI) device used to drive the ECDL current. This leads to realize a servo loop which compensates the fast frequency fluctuations (Fig. 6). In a free running operation, the width of the optical beatnote (in the green range) between the two IR lasers was estimated at a few hundreds of kHz mainly due to the linewidth of the ECDL. It has been reduced to 30 kHz thanks to the used phase loop setup developed for this work.

The phase lock bandwidth is about 450 kHz as shown in Figure 7, yielding to about 78% of the measured spectral power contained in the carrier of the beatnote in phase-locked conditions. The rest of the power resides in the noise pedestal and in the servo bumps observed on either side of the carrier.

In conclusion, we have presented the phase locking of two infrared lasers emitting at 1029 nm and 1544 nm, using doubling and tripling processes respectively. With this preliminary scheme, we report 1.5 mW of green power generated at 515 nm from 630 mW at 1544 nm with frequency tripling process. Furthermore, we produce more than 30 mW in the green from a frequency doubled 1029 nm source. The generation of the 515 nm radiations opens the way to the use of very narrow iodine lines for the frequency stabilisation purpose of our MFC.

The phase locking of both IR sources leads to a set of four phase coherent radiations at 1544 nm, 1029 nm, 772 nm and 515 nm. The obtained output powers at these wavelengths are at least 3 orders of magnitude higher than the available power from each mode of currently used frequency combs.

**Acknowledgments**
This work is supported by Observatoire de Paris, GRAM (CNRS/INSU/CNES), DGA (ANR N° 11 ASTR 001 01) and LNE. N.C. is grateful to the Mairie de Paris for financial grant via « the Research in Paris 2010' program ».
J. H. acknowledges the support of GACR (GPP102/11/P820), MYES (CZ.1.05/2.1.00/01.0017), European Social Fund and National Budget of the Czech Republic (CZ.1.07/2.4.00/31.0016).

# Figures

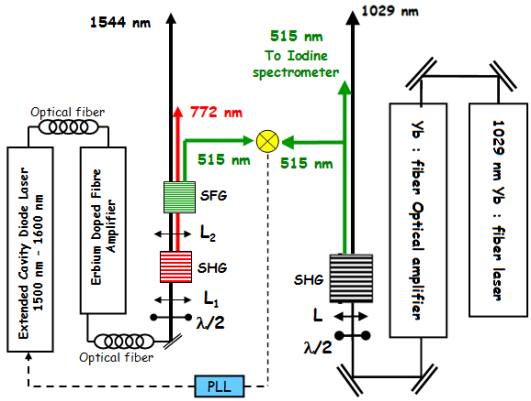

Fig. 1: Schematic of the whole optical setup.

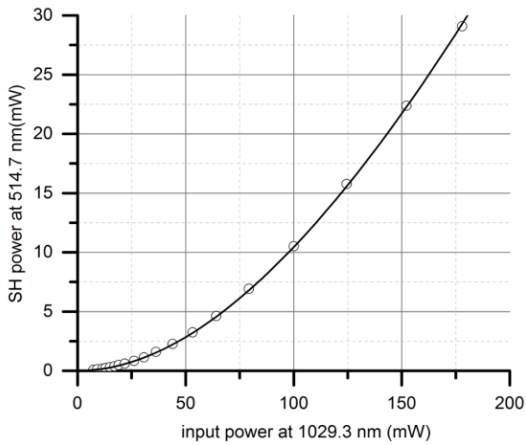

Fig. 2: SH power versus IR power collected at fixed crystal temperature (66 °C) and fibre laser wavelength (1029 nm). The solid curve corresponds to a fit using $P_{2\omega} = \tau P_\omega \tanh^2\left(\sqrt{\eta L^2 \tau P_\omega}\right)$, where $\tau = 45\%$ is the coupling efficiency at the waveguide input, $L$ = 1 cm is the waveguide crystal length, and $\eta = 613$ %W$^{-1}$ cm$^{-2}$ is the normalized conversion efficiency determined from low power measurements [10].

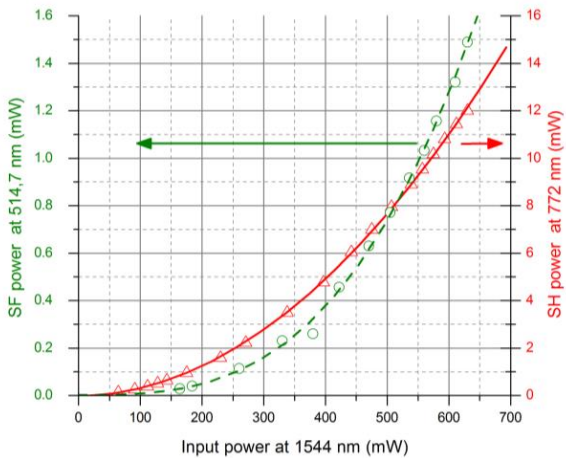

Fig. 3: Harmonic power generated at 2ω (red curve) and 3ω (green curve) versus IR power at ω (1544 nm). The red solid curve is a quadratic fit for the SHG process, and the green dashed curve is a cubic fit for the SFG process.

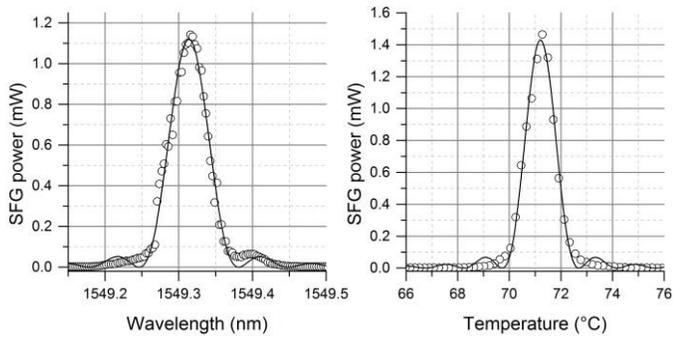

Fig. 4: Harmonic power at 516 nm versus IR power.
(a) THG power versus the fundamental input wavelength.
(b) THG power versus the crystal temperature.
The solid lines are a sinc$^2$ fit of experimental data.

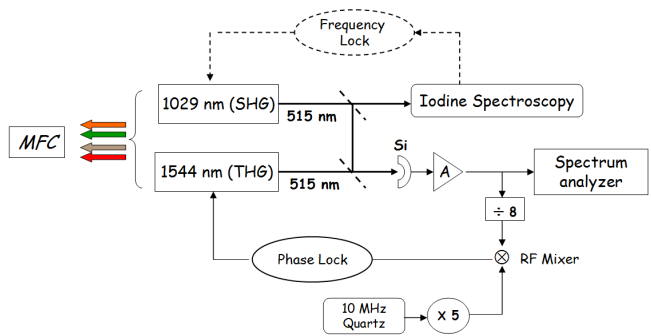

Fig. 5: Phase lock setup of the MFC.

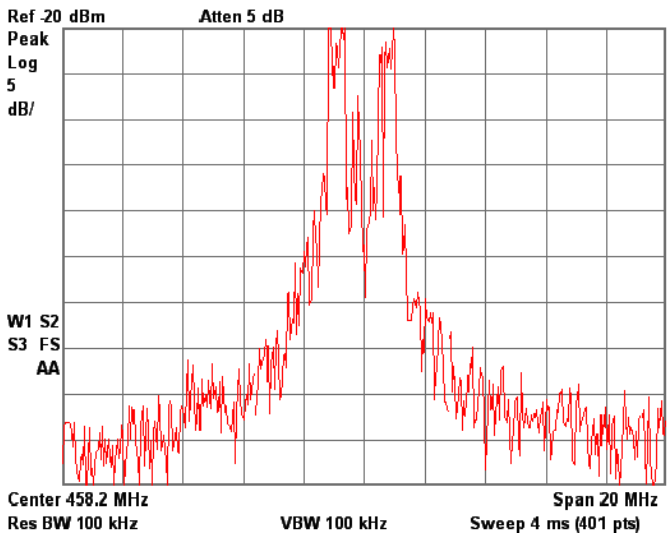

Fig. 6: RF beat notes at 515 nm. The two IR lasers operate in free-running regime (RBW = 100 kHz).

Fig. 7: The ECDL is phase locked against fibre laser with 500 kHz bandwidth (RBW = 10 kHz).